\documentclass[conference,review,anonymous]{IEEEtran}
\IEEEoverridecommandlockouts
\usepackage{amsmath,amsfonts}
\usepackage{textcomp}
\usepackage{xcolor}
\usepackage{algpseudocode}
\usepackage{array}
\usepackage{enumitem}
\usepackage{booktabs}
\usepackage{caption}
\usepackage{cmap}
\usepackage{colortbl}
\usepackage{comment}
\usepackage{datatool}
\usepackage{epstopdf}
\usepackage{etex}
\usepackage{fancyvrb}
\usepackage{fixltx2e}
\usepackage{filecontents}
\usepackage{float}
\usepackage{framed}
\usepackage{fp}
\usepackage{graphicx}
\usepackage{hhline}
\usepackage{hyperref}
\usepackage[utf8]{inputenc}
\usepackage{lineno}
\usepackage{listings}
\usepackage{longtable}
\usepackage{makecell}
\usepackage{multirow}
\usepackage{pdfpages}
\usepackage{pgf}
\usepackage{pgfplots}
\usetikzlibrary{patterns}
\usepgfplotslibrary{statistics}
\usepackage{pgf-pie}
\usepackage{soul}
\usepackage{subfig}
\usepackage{tabularx}
\usepackage{tikz}
\usepackage{tcolorbox}
\usepackage[normalem]{ulem}
\usepackage{url} 
\usepackage{xpatch}
\usepackage{tikz}
\usetikzlibrary{fadings}

\pgfplotsset{compat=1.11}
\definecolor{dkgreen}{rgb}{0,0.6,0}
\definecolor{gray}{rgb}{0.5,0.5,0.5}
\definecolor{mauve}{rgb}{0.58,0,0.82}
\definecolor{lgray}{rgb}{0.98,0.98,0.995}
\definecolor{darkslategray}{rgb}{0.18, 0.31, 0.31}
\definecolor{darksienna}{rgb}{0.24, 0.08, 0.08}
\definecolor{dimgray}{rgb}{0.41, 0.41, 0.41}
\definecolor{Gray}{gray}{0.9}

\makeatletter
\def\setcolor#1\pgfeov{\def\pgfpie@color{#1}}
\pgfkeyslet{/color/.@cmd}{\setcolor}
\xpatchcmd{\pgfpie@findColor}{\color}{\pgfpie@color}{}{}
\xpatchcmd{\pie}{\color}{\pgfpie@color}{}{}
\makeatother

\newlength{\maxlen}

\def\BibTeX{{\rm B\kern-.05em{\sc i\kern-.025em b}\kern-.08em
    T\kern-.1667em\lower.7ex\hbox{E}\kern-.125emX}}
\begin{document}

\title{Annotating Privacy Policies in the Sharing Economy}

\author{\IEEEauthorblockN{Fahimeh Ebrahimi, Miroslav Tushev, and Anas Mahmoud}
\IEEEauthorblockA{\textit{The Division of Computer Science and Engineering} \\
\textit{Louisiana State University}\\
febrah1@lsu.edu, mtushe1@lsu.edu, amahmo4@lsu.edu}}

\maketitle

\begin{abstract}
Applications (apps) of the Digital Sharing Economy (DSE), such as Uber, Airbnb, and TaskRabbit, have become a main enabler of economic growth and shared prosperity in modern-day societies. However, the complex exchange of goods, services, and data that takes place over these apps frequently puts their end-users' privacy at risk. Privacy policies of DSE apps are provided to disclose how private user data is being collected and handled. However, in reality, such policies are verbose and difficult to understand, leaving DSE users vulnerable to privacy intrusive practices. To address these concerns, in this paper, we propose an automated approach for annotating privacy policies in the DSE market. Our approach identifies data collection claims in these policies and maps them to the quality features of their apps. Visual and textual annotations are then used to further explain and justify these claims. The proposed approach is evaluated with 18 DSE app users. The results show that annotating privacy policies can significantly enhance their comprehensibility to the average DSE user. Our findings are intended to help DSE app developers to draft more comprehensible privacy policies as well as help their end-users to make more informed decisions in one of the fastest growing software ecosystems in the world.
\end{abstract}


\section{Introduction}
The Digital Sharing Economy (DSE), also known as the \textit{gig} or \textit{shared} economy, refers to a sustainable form of business exchange that provides access to, rather than ownership of, assets and resources via direct Peer-to-Peer (P2P) coordination~\cite{Martin16}. Over the past decade, applications of the Sharing Economy, such as Uber, TaskRabbit, and Airbnb, have caused major disturbances in established classical markets, enabling people to exchange and monetize their idle or underused assets at unprecedented scales. This unique form of collaborative consumption has been linked to significant levels of economic growth, helping unemployed and partially employed individuals to generate income, increase reciprocity, and access resources that can be unattainable otherwise~\cite{Dillahunt15,Williams20,Hamari16}. As of today, there are thousands of active DSE platforms, operating in a market sector that is projected to grow to close to 335 billion U.S. dollars by 2025~\cite{Se15}. 

In order to mediate business transactions, DSE apps constantly demand access to private user information, including their credit information, geo-location, and even photos of the assets being shared (e.g., a vehicle or an apartment)~\cite{Lutz18,Ranzini17,Mare20,Pham17}. Such information is used to establish the P2P connection between service providers and receivers, facilitate identification offline, and optimize and manage transactions. Modern app stores require apps that operate on users' sensitive information to provide privacy policies in which their data practices are declared and justified~\cite{Bhatia16Mining}. These policies are intended to provide information about the types of user information the app collects, how that data is being used, shared, transferred, and protected~\cite{Bhatia18,Keymanesh20}. However, in reality, these policies are often verbose, ambiguous, and jargon-heavy, making them difficult to understand by the average user~\cite{Wang18,Bhatia19}. This can be particularly problematic in the DSE market, where an unintended leak of users' private information can expose them to great physical and financial risks~\cite{Ranzini17,Mare20,Pham17}.

To help overcome these limitations, in this paper, we propose a novel approach for annotating privacy policies in the DSE market~\cite{Wilson16}. Our objective is to make these policies more comprehensible to the average DSE user. Providing easy-to-understand policies can protect users from the intrusive privacy tactics of apps and help them make more informed socio-economic decisions when it comes to navigating the landscape of DSE platforms. Technically, our approach utilizes text classification techniques to extract data collection practices from DSE privacy policies and then map these practices (claims) to the quality features of their apps. The extracted information is then color-coded, annotated, and presented using an enhanced hypertext format. The impact of our annotations is then evaluated through a user study with 18 DSE app users.

The remainder of this paper is organized as follows. Section~\ref{sec:bg} discusses privacy in the DSE market and motivates our research. Section~\ref{sec:approach} describes our data collection and analysis process. Section~\ref{sec:automated} presents our automated policy annotation approach. Section~\ref{sec:eval} empirically evaluates our proposed approach. Section~\ref{sec:discuss} discusses our findings and their implications. Section~\ref{sec:threats} addresses the limitations of our study. Finally, Section~\ref{sec:conclusion} concludes the paper.

\section{Background and Motivation}
\label{sec:bg}
In this section, we discuss user privacy in the DSE market and review existing research on privacy policy annotation. We then motivate our work and present our research questions.

\subsection{Privacy in the Digital Sharing Economy}
\begin{figure}\centering
\includegraphics[scale=0.65]{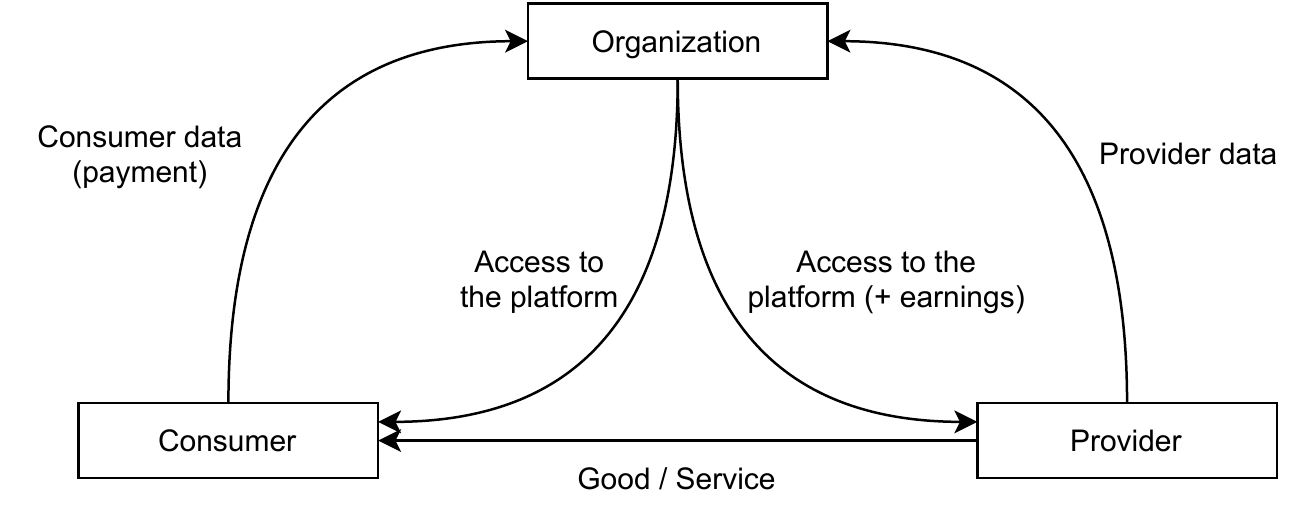}
\caption{A summary of data exchange between users and organizations in DSE apps~\cite{Ranzini17}.}
\label{fig:privacyinSE}
\end{figure}

The proliferation of DSE apps over the past decade along with their unique operational characteristics have imposed new challenges on end-users privacy. Privacy in the DSE market is a compounded concept, including interrelated concerns of data and physical privacy~\cite{Lutz18}. Fig.~\ref{fig:privacyinSE} shows the different channels of data exchange in a typical DSE transaction~\cite{Ranzini17}. In order to be approved as a service provider (e.g., Uber drivers or Airbnb hosts), users need to disclose their personally identifiable information (PII) to DSE organizations (e.g., Uber or Airbnb), including their legal name, address, and picture along with their shared property's information, such as the model of the vehicle being shared or the number of rooms in the rental space. Service consumers (e.g. Uber riders or Airbnb renters) also need to share their PII and financial information with DSE companies in order for their service requests to be approved. DSE companies use their users' information to mediate their transactions (e.g., route the Uber driver) and charge for the service. Some of this information is also made available to service providers and receivers to facilitate identification offline. %for example, for an Airbnb host to confirm the identity of their guests or for a rider to identify their Uber driver.    

Psychical privacy in the Sharing Economy refers to the threats that are often associated with sharing personal resources with strangers~\cite{Mao20,Pham17}. Such concerns tend to be domain-specific, influenced by the nature of the DSE transaction. For instance, recent research has revealed that the presence of smart devices in Airbnb rentals can raise privacy concerns among guests, including concerns of excessive monitoring and access to their Internet history~\cite{Mare20}. A survey of Uber riders exposed common concerns of stalking, coercion, and habit discovery among both riders and drivers~\cite{Pham17}. DSE platforms attempt to mediate these concerns by establishing trust in the process. Trust is a key user goal in DSE~\cite{Huurne17,Ert16,Mao20,Carin19}. Providers and receivers at both ends of the P2P connection need to maintain a minimum level of mutual trust before a transaction can take place~\cite{Carin19,Ert16}. However, in order to establish trust, apps demand the disclosure of even more personal information (e.g. social media accounts or hobbies) to make their users appear more trustworthy~\cite{Ma17,Picard18}.  

In general, privacy in the DSE market is a multidimensional problem. The problem is further exacerbated by the fact that DSE apps extend over a broad range of application domains~\cite{Tushev21}. Identifying and mitigating privacy threats in each specific domain can be a very challenging task, and the failure to do so can lead to a decline in sharing intensity~\cite{Hayes18}. This can prevent users, especially in communities at the lower end of the economic ladder, from reaching their full economic and social potential in the DSE market~\cite{Ranzini17}.

\subsection{Privacy Policy Annotation}
Privacy policies are unilateral contracts by which organizations describe their data practices and inform end-users about their data collection, usage, and sharing practices. Popular app marketplaces, such as Google Play and the Apple App Store, require apps to post their privacy policies online. However, recent analysis of privacy policies in the mobile app market revealed that most of these policies are ambiguous, and oftentimes contain incomplete or deceptive information~\cite{Reidenberg16}. This undermines the utility of these policies and diminishes users' trust in their apps~\cite{Bhatia16}. 

Several methods have been proposed in the literature to enhance the quality of privacy policies. Such techniques commonly rely on annotating the content of these policies with supplemental information to clarify ambiguous privacy claims and justify data usage practices. Annotations are often carried out manually, either by domain experts or using crowd-sourcing~\cite{Breaux14,Sadeh13,Wilson16Crwod,Bhatia19,Bhatia16}. Expert annotations, while usually accurate, are frequently associated with significant effort. For instance, forming multidisciplinary expert teams for specific application domains can be a time-consuming process that hardly scales up. Crowd-sourcing can also lead to suboptimal results due to the general lack of technical and legal knowledge necessary to produce correct annotations~\cite{Wilson16,Reidenberg15,Reidenberg16,Sadeh13,Wilson16Crwod}.  

To overcome these limitations, several attempts have been made to automate the annotation process~\cite{Wilson16}. For instance, Wilson et al.~\cite{Wilson16} annotated a corpus of 128 policies to 23,000 fine-grained data practices to be used as a benchmark in automated annotation tasks. In general, automated techniques use Natural Language Processing (NLP) and Machine Learning (ML) to identify salient paragraphs that describe data practices, such as data collection, sharing, security, and retention in privacy policies~\cite{Wilson16,Ramanath14}. The main limitation of such methods is that they are commonly applied as generic solutions. This can impact the validity of these techniques due to the fact that privacy concerns tend to be domain-specific. For instance, constant location tracking is more justifiable in ride-hailing apps than in social media apps. Therefore, a \textit{one-size-fits-all} solution often generates misleading results.

\subsection{Motivation and Research Questions}
Our brief review shows that people's participation in sharing activities might be hindered by the unjustified privacy practices of DSE apps. In general, people might abstain from sharing if they cannot justify the trade-off between the benefits and the risks of using DSE apps. DSE apps try to mitigate these concerns in their privacy policies. However, the majority of these policies are extremely lengthy, and oftentimes ambiguous, leaving users exposed to unlawful privacy practices, such as constantly and unnecessarily tracking their location, learning their routine habits, and inferring their consumption preferences~\cite{Pham17,Mare20}. For instance, the average length of privacy policies of popular DSE apps (Table~\ref{Tab:dataset}) is around 5,425 words, which is significantly longer than the average length of online privacy policies (2,000 $\sim$ 3,000 words)~\cite{Waldman18,Liu21,Meier20}. Furthermore, the average readability of these apps' policies, as measured by the Flesch Reading Ease (FRE)~\cite{Flesch49,Mcdonald09,Powell18}, is around 29, which indicates that they are better understood by college graduates. This can be particularly problematic in the DSE market as recent statistics showed that a considerable percentage of DSE users either do not possess a college-level education or still college students~\cite{Hsiao18,Se15}. These observations emphasize the need for a supporting mechanism to help DSE users better assess the privacy risks of disclosing their personal information to DSE apps. In fact, such a mechanism can also help policymakers and regulators to better evaluate the privacy practices of DSE apps, and consequently, devise legislation to protect workers' and consumers' rights in one of the fastest-growing, most diverse, yet under-regulated software ecosystems in the world~\cite{Rogers15,Leong17}.

Motivated by these observations, in this paper, we propose a novel approach for annotating privacy policies in the DSE market. Our approach automatically classifies data claims in these policies and maps them to the quality features of the app (e.g., security, safety, customizability, etc.). The main assumption is that these generic categories of abstract system features can be more easily comprehensible for the average user. In particular, our research questions are:    

\begin{itemize} 
\item \textit{\textbf{RQ$_1$:} Can the privacy policies of DSE apps be automatically annotated?} Under this research question, we investigate the effectiveness of several automated classification techniques in annotating data collection claims in the privacy policies of DSE apps. 
\item \textit{\textbf{RQ$_2$:} Are annotated privacy policies more comprehensible to the average DSE user?} Under this research question, we explore whether our annotated policies can actually help average DSE users to better understand the privacy practices of their DSE apps.  
\end{itemize}

\section{Data and Manual Annotation}
\label{sec:approach}
\begin{figure*}[ht]\centering
\includegraphics[trim=2.9cm 10.5cm 1.9cm 3.3cm, clip=true,scale=0.55]{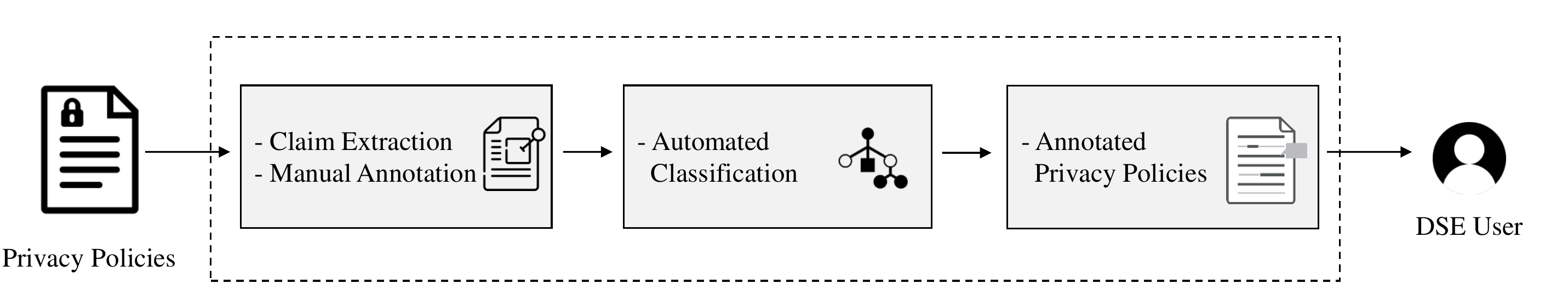}
\caption{A diagram of the proposed approach.}
\label{fig:approach}
\end{figure*}

Our approach can be divided into four main steps, policy collection, manual annotation, automated classification, and policy presentation. A summary of the approach is presented in Fig.~\ref{fig:approach}. In what follows, we describe our data collection and manual annotation steps.  

\subsection{Policy Collection}
Recent statistics estimate that there are hundreds of active DSE platforms listed on popular mobile app marketplaces~\cite{Tushev21}. In our analysis, we consider apps that operate in large geographical areas and have massive user bases. Privacy concerns are more likely to manifest over these apps rather than smaller ones which often have less heterogeneous user bases. Specifically, for a DSE app to be included in our analysis, it has to meet the following criteria:

\begin{enumerate}
\item The app must facilitate some sort of a P2P connection and include the sharing of some sort of a resource, such as a tangible asset (e.g., an apartment or a car) or a soft skill (e.g., plumbing or hair styling). 

\item The app must be available on Google Play or the Apple App Store so that we can extract its meta-data. 

\item The app must be located and/or have a substantial presence in the United States. By focusing on the U.S. market, we ensure that our apps' privacy policies are available in English and that these apps offer services that are familiar to the average U.S. user. 
\end{enumerate}

With these criteria in place, we selected the five most popular apps from five popular application domains of DSE~\cite{Tushev21}. 
In general, five categories of DSE apps can be identified: ride-sharing (e.g., Uber or Lyft), lodging (e.g., Airbnb), delivery (e.g., DoorDash or UberEats), asset-sharing (e.g., GetMyBoat), and freelancing (e.g., TaskRabbit)~\cite{Tushev21, Ebrahimi21}. The top five apps in each application domain are then identified based on their installation and rating statistics as of January, 2021. Table~\ref{Tab:dataset} shows the selected apps along with their popularity, measured as the number of ratings and the average rating on the Apple App Store as well as the average number of installs from Google Play. We also extracted each app's privacy policy, which is typically posted on the app's official website. 

%dataset
\begin{table*}
\footnotesize 
\centering
\renewcommand{\arraystretch}{1.2}
\caption{The DSE apps included in our analysis along with their application domains and measures of popularity.}
\label{Tab:dataset}
\smallskip 
\begin{tabular*}{55em}{l|l|c|c|c}
\Xhline{1.5\arrayrulewidth}
\textbf{Domain} & \textbf{Apps} & \textbf{Avg. Rating} & \textbf{Avg. \# Reviews} & \textbf{Avg. \# Installs} \\ \Xhline{1.5\arrayrulewidth}
Ride-hailing & Uber, Lyft, Via, Gett, Curb, Scoop & 4.7 & 1.7M & 87M \\\hline
Lodging &  Airbnb, Vrbo, Dyrt, Hostelworld, HomeToGo, Booking & 4.8 & 322K & 26M\\\hline
Delivery &  UberEats, DoorDash, Instacart, GrubHub, Postmates, goPuff & 4.7 & 1.3M & 22M \\\hline
Skill-based & TaskRabbit, Fiverr, Thumbtack, Upwork, Handy, StyleSeat & 4.7 & 98K & 2.5M \\\hline
Asset Sharing &  Turo, Getaround, GetMyBoat, HyreCar, Outdoorsy, Avail & 4.7 & 31K & 285K \\
\Xhline{1.5\arrayrulewidth}
\end{tabular*}
\end{table*}

\subsection{Manual Annotation}
We start our analysis by qualitatively analyzing the content of privacy policies of the apps in our dataset. The objective is to identify the data collection claims in these policies along with their justifications (i.e., establish our ground truth). We define a justification as the rationale provided by the app for collecting users' sensitive information. In our approach, we map such rationale into a set of high-level system quality features. Apps supposedly collect data to enhance the quality attributes of the app, and thus, its users' experience. These quality attributes are often described as the non-functional requirements of the system (NFRs). NFRs can be thought of as abstract behaviors of the system that can be enforced through bundles of the system's functional features. For instance, the \textit{security} NFR refers to the behavior that is enforced by the functional features that are used to implement security in the system, such as user authentication and data encryption.  

Around 250 different kinds of software quality attributes are defined in the literature~\cite{Mairiza10}. These NFRs extend over a broad range of categories and sub-categories. To simplify our manual analysis, we limit our annotation to the most popular types of NFRs that commonly appear in literature: Security, Performance, Accessibility, Accuracy, Usability, Safety, Legal, and Maintainability~\cite{Cleland07,Slankas13,Jha19,Mahmoud16}. These NFRs are defined in Table~\ref{Tab:examples}. 

\begin{figure}\centering
\includegraphics[clip=true,scale=0.6]{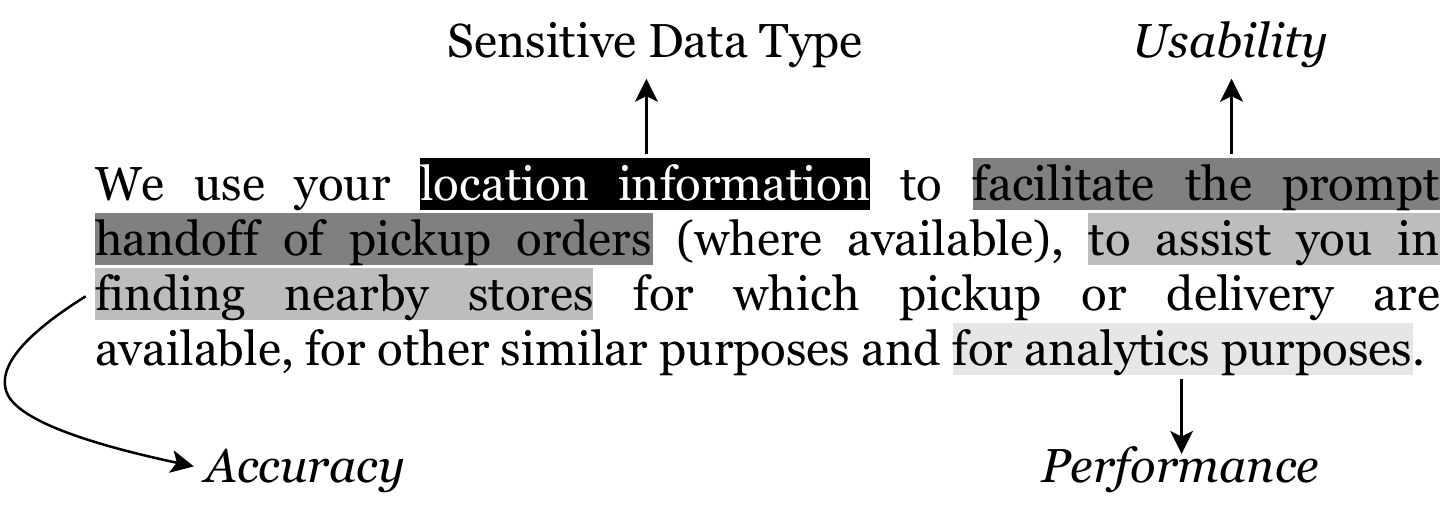}
\caption{An example of annotating a privacy statement.}
\label{fig:annotateExample}
\end{figure}

%Privacy policies often include multiple parts, such as data collection, data usage, data sharing, and children's privacy. To avoid information overload, in our analysis, we only annotate the data collection, usage, and sharing of personal information part~\cite{Bhatia19,Anton04}.
To annotate the privacy claims in our set of policies, we follow a grounded theory approach~\cite{Corbin98}. In particular, three judges manually extracted any policy statements related to collecting, using, sharing, or storing personal information and mapped these statements to one or more of the NFR categories defined earlier. If no suitable category was found, the judges were free to come up with new categories. An example of this process is shown in Fig.~\ref{fig:annotateExample}. A statement can be labeled under multiple categories if it raises more than one functionality-related issue. This step was necessary to maintain the accuracy of our annotations as NFRs are inherently vague—a single statement can express multiple issues at the same time~\cite{Jha19,Mairiza10,Glinz07}. After each round of annotation, the three judges met to discuss any discrepancies and add/merge labels. This process was continued until the judges reached a consensus for each annotated policy. 

%For example, the statement ``\textit{help you to use our services faster and more easily through features such as the ability to sign-in using your account}'' is classified under \textit{performance} and \textit{usability}. 

The outcome of our manual annotation process is shown in Table~\ref{Tab:examples}. In total, 639 statements were extracted and classified under 11 different types of NFRs. Table~\ref{Tab:examples} shows these NFRs, their definitions, an example statement of each type, the number of apps in our dataset that include each NFR in their privacy policy, and the number of statements classified under each NFR. In addition to the 11 general categories of NFRs, two new types of NFRs emerged during our classification process: Trust and Customizability. As mentioned earlier, trust is a key user goal in the DSE market. Such goals frequently appeared in statements referring to enhancing the apps' trustworthiness by anonymizing personal information, verifying users' identities, and publicizing profile information. Customizability includes any statement related to providing personalized services to meet users' preferences, for example ``\textit{personalize and customize your experience based on your interactions with the Airbnb Platform''}.

%\footnote{https://github.com/SEGroup-2318/ICSE}
%Examples of such statements include, ``\textit{Uber uses upfront pricing to let riders know the cost of their trip before they request a ride}'' and ``\textit{your profile information is public to allow our community to evaluate your reliability and responsiveness''}. 

Out of the 639 identified statements, 34.42\% (220) were mapped into more than one NFR category. Usability (19.71\%) and Performance (16.58\%) were the two most frequently used NFRs, followed by Security (14.86\%) and Legal (12.2\%). The privacy policies of all of the 30 DSE apps examined included statements related to Usability, Performance, and Security. Accessibility (0.007\%) is the least common issue. Only three apps (Uber, UberEats, and HyreCar) used Accessibility to justify personal data collection, for example, ``\textit{enable Accessibility features that make it easier for users with disabilities to use our services}''. Out of the 639 statements annotated, 39 statements did not fit under any NFR category. For example, several apps claimed that they collect personal data for research purposes (e.g., ``\textit{conducting research and analysis}''). Such statements were labeled as \textit{Other}.

%It in important to point out that some NFR categories are more well-defined (i.e. Security and Performance) than others. Policy statements can be easily mapped to these categories. However, some other cases are less well-defined. For instance, Usability includes statements related to the ease of use, look and feel, consistency, attractiveness, and overall layout of the app. Similarly, Maintainability statements often describe issues related to the reliability and availability of the app under normal and abnormal conditions. This makes the process of classifying statements under these categories more challenging. 
%Examples of such statements include ``\textit{ensuring the comfortable use of our website/application}'' and ``\textit{provide customer support and contact you from time to time}''.  Examples of such statements include ``\textit{we will use your data in order to fix any bugs that we find}'' and ``\textit{ensuring that a trouble-free connection is established}''. 

\begin{table*}
\footnotesize 
\centering
\renewcommand{\arraystretch}{1.2}
\caption{The set of NFRs used in our manual annotation. The table also provides examples on each type, the number of apps each NFR appears in, and the number data collection statements classified under each NFR.}
\label{Tab:examples}
\smallskip 
\begin{tabular*}{\textwidth}{p{0.09\linewidth} | p{0.365\linewidth} | p{0.305\linewidth} | c | c }
\Xhline{1.5\arrayrulewidth}
\textbf{NFR} & \textbf{Description}& \textbf{Example Statement} & \textbf{\#Apps} & \textbf{\#Statements} \\ \Xhline{1.5\arrayrulewidth}

Usability & Any requirement that specify the end-user-interactions with the system and the effort required to learn, operate, prepare input, and interpret any system outputs & ``\textit{Simplify your access to and use of the Platform and make it more seamless}''& 30 & 126\\
\hline

Performance & Any requirement that specifies the capability of a software product to provide appropriate performance relative to the number of resources needed to perform effectively under stated conditions & ``\textit{Matching available drivers and delivery persons to users requesting services, aiming to reduce the average wait time for everyone}'' & 30 & 106 \\
\hline

Security & Any requirement which prevents unauthorized access to the system, programs, and data & ``\textit{Detect and prevent fraud, spam, abuse, security and safety incidents, and other harmful activity}'' &  30 & 95\\
\hline

Legal & Any requirement that specifies the software capability to guarantee users rights to access/delete their personal data, and resolve their legal complaints against the platform  &  ``\textit{In certain jurisdictions, you can request that your personal information be deleted}''& 29 & 78\\
\hline

Safety & Any requirement that specifies the software capability to ensure the state of being "safe", the condition of being protected from harm or other non-desirable outcomes & ``\textit{We use data from drivers devices to help identify unsafe driving behavior such as speeding or harsh braking and acceleration}''& 25 & 75\\
\hline	

Customizability & Any requirement that specifies the capability of a software product to personalize and customize services according to its users' preferences &``\textit{Personalize and customize your experience based on your interactions with the Airbnb Platform}''&  26 & 52  \\
\hline

Accuracy &  Any requirement that is concerned with defining the precision which the system records or produces data. &``\textit{Share information with Google, in connection with the use of Google Maps in Uber apps}'' & 18 & 27\\
\hline

Maintainability & Any requirement that specifies the capability of a software product to operate without failure and maintain a certain level of performance when used under normal conditions during a given time period &  ``\textit{Perform internal operations necessary to provide our services, including to troubleshoot software bugs and operational problems}''&  18 & 20\\
\hline

Trust & Any requirement that is used to enhance users' confidence, faith, or hope in the software system &``\textit{Uber uses upfront pricing to let riders know the cost of their trip before they request a ride to give them peace of mind}''& 11 & 16 \\
\hline

Accessibility & Any requirement that specifies the capability of a software product to be accessible and usable by all users, including people with disabilities &``\textit{Enable Accessibility features that make it easier for users with disabilities to use our services}''& 3 & 5 
 \\
 \hline

Other &  Statements that specify the capability of the software to provide services, enable users to refer friends, or the system to conduct research, are included in this category &``\textit{we will use your data to provide you with content, products, and services you request
}'' &  23 & 39  \\ \hline

\Xhline{1.5\arrayrulewidth}
\end{tabular*}
\end{table*}

\section{Automated Annotation}
\label{sec:automated}
Under this step, we investigate the performance of automated classification techniques in classifying the data collection statements in DSE privacy policies into the different NFR categories identified earlier. This process can be described as a multi-labeling classification problem~\cite{Ghamrawi05}. One commonly used approach to perform multi-label classification is Binary Relevance (BR). BR assumes label independence. Specifically, it decomposes a problem with $n$ classes (labels) into $n$ binary problems. In each problem, a single label $i$ is considered to be the correct class and the rest of classes $(n - 1)$ are considered to be incorrect. BR then learns a single binary model for each of the $n$ binary problems~\cite{Tsoumakas09}. The output of the classifier is the union of predictions, where a data sample is assigned to any label $i$ if the classifier classified it under that label. %In what follows, we describe our classification process in detail. 

%Despite its sometimes unrealistic label dependence assumption, BR has been successfully used in tasks such as assigning keywords to scientific papers, illnesses to patients, and emotional expressions to human faces~\cite{Luaces12}.  Our classification configurations can be described as follows: 

%, where each statement can be classified under one or more labels.
%Formally, let $L$ be a finite and non-empty set of labels ${l_1, ..., l_L}$, let $Y$ be an input space, and let $Z$ be the output space defined as a subset of the set of labels $L$. A multi-label classification task can be given by $D = (y_1, z_1), ..., (y_n, z_n) \subset Y \times Z$, where $(y_1, z_1)$ is the classification of the data instance $y_1 \in Y$ under the label $z_1 \in Z$. 
%In multi-label classification, a data sample can be classified under one or more labels.

\subsection{Classification settings}
To classify our data, we experiment with Support Vector Machines (SVM)~\cite{Joachims98}. This algorithm is commonly used to classify online text~\cite{Guzman15,Maalej15, Williams20}. Its success can be attributed to its ability to deal with short texts (e.g., tweets, user reviews, youtube comments, etc.)~\cite{Wang12,Poche17}. Our analysis is performed using Scikit-learn, a Python library that integrates a wide range of state-of-the-art Machine Learning algorithms for supervised classification problems~\cite{Pedregosa11}. 

Combinations of text pre-processing strategies are often used in text classification tasks to remove potential noise and to enhance the prediction capabilities of the classifier~\cite{Joachims98}. In our analysis, policy statements were first converted into lower case tokens. Tokens that contained non-ASCII characters, digits, and URLs, were removed. English stop-words (e.g., \textit{the, in, will}) were also removed based on the list of stop-words provided in NLTK~\cite{Loper02}. The remaining words were then lemmatized. We selected lemmatization over stemming to preserve the naturalness of words. 

To represent each statement in our policies we used two different methods, Vector Space Model (VSM) and the word embedding algorithm GloVe. VSM is an algebraic model that consists of a single term-document matrix. Each entry in the matrix $w_{i,j}$ is the weight of the term $j$ in the document $i$, indicating the importance of the term to the document's subject matter. Term frequency-inverse document frequency (TF.IDF) is used to assign a weight for each word proportional to its importance to the text~\cite{schutze08}. TF.IDF is calculated as the product of the frequency of the term in the document (TF) and the term's scarcity across all the documents (IDF). 

GloVe~\cite{Pennington14} is a popular word embedding that uses the similarities between words as an invariant to generate their vector representations. In general, word embeddings represent individual words in a corpus using multi-dimensional vectors of numeric values that are derived from the intrinsic statistical properties of the corpus. GloVe initially constructs a high dimensional matrix of words co-occurrence. Dimensionality reduction is then applied to the co-occurrence count matrix of the corpus. By applying a matrix factorization method on the count matrix, a lower dimension matrix is produced, where each row is the vector representation of a word. To conduct our analysis, we converted the list of pre-processed tokens in each policy statement into a vector of word embeddings using the pre-trained model of GloVe. We then used the generated word embeddings to represent the statement. Word collection (e.g., phrase, sentence, or paragraph) embeddings can be computed using operations on word vectors, such as their unweighted averaging/summation~\cite{Mikolov13-2}, Smooth Inverse Frequency (SIF)~\cite{Arora16}, and Doc2Vec~\cite{Le14,Safadi20}. In our analysis, we used the simple unweighted averaging method to obtain an embedding for each statement in apps privacy policies~\cite{Arora16,Sadeghian15}. Averaging word vectors has been proven to be a strong baseline for paragraph representation, especially in cases when the order of words in the text is unimportant~\cite{Kenter16}. 

To train and test our classifiers, we use \textit{10}-fold cross-validation. This approach creates 10 partitions of the dataset. In each partition, 90\% of the instances are considered as the training set and 10\% as the test set. 10-fold cross-validation is selected over other techniques, such as the holdout method (e.g. train/test split), to decrease the variance of the results.

The outcome of multi-label classification can be either fully correct, partially correct, or completely incorrect. Therefore, the standard precision and recall metrics, typically used in binary classification tasks, need to be accompanied by other measures that can account for partial correctness. To account for such information, we use subset accuracy (SA), hamming score (HS), and hamming loss (HL)~\cite{Godbole04}. SA is the number of predictions that are completely correct divided by the total number of classified data instances. HS is the proportion of correctly predicted labels over the total number of labels identified for a data instance. HL is a measure of the number of times on average a class label is incorrectly predicted. %In addition, we compute the standard measures of macro-averaged Precision (P), Recall (R), and F-Measure ($F_2$) for each classification label and averaged over all the labels.

%In particular, stemmers (e.g., Porter stemmer~\cite{Porter80}) tend to be prone to over-stemming which happens when too much of the word is removed that the outcome is not a valid natural word (e.g., \textit{general} and \textit{generous} are stemmed to \textit{gener}). This can be a key factor in the performance of methods that use English corpora for similarity calculations.

%We selected the averaging instead of summation to handle different length sentences.
\begin{table*}
\footnotesize 
\centering
\renewcommand{\arraystretch}{1.2}
\caption{The performance of Binary Relevance (BR) classification algorithm under our different classification configurations (VSM: Vector Space Model)}
\label{Tab:results}
\smallskip 
\begin{tabular*}{48.5em}{l|c|c|c|c|c|c|c|c|c|c|c|c}
\Xhline{1.5\arrayrulewidth}
\multirow{2}{*}{\textbf{Functionality}} &  \multicolumn{6}{c|}{\textbf{VSM}} & \multicolumn{6}{c}{\textbf{GloVe}} \\ \cline{2-13}

 & \textbf{P} & \textbf{R} & \textbf{$F_2$} & \textbf{HS} & \textbf{SA} & \textbf{HL} & \textbf{P} & \textbf{R} & \textbf{$F_2$} & \textbf{HS} & \textbf{SA} & \textbf{HL} \\\Xhline{1.5\arrayrulewidth}

Usability        	& 0.75	& 0.57	& 0.59	& 0.86	& 0.83	& 0.09	& 0.86	& 0.74	& 0.76	& 0.92	& 0.89	& 0.09 \\ \hline
Performance      	& 0.58	& 0.53	& 0.53	& 0.79	& 0.72	& 0.17	& 0.84	& 0.76	& 0.77	& 0.85	& 0.81	& 0.14 \\ \hline
Security         	& 0.7	& 0.56	& 0.58	& 0.8	& 0.74	& 0.15	& 0.85	& 0.78	& 0.79	& 0.86	& 0.84	& 0.13 \\ \hline
Legal            	& 0.87	& 0.69	& 0.71	& 0.86	& 0.81	& 0.08	& 0.99	& 0.93	& 0.94	& 0.94	& 0.91	& 0.05 \\ \hline
Safety           	& 0.64	& 0.51	& 0.53	& 0.83	& 0.77	& 0.12	& 0.79	& 0.65	& 0.67	& 0.88	& 0.82	& 0.11 \\ \hline
Customizability  	& 0.6	& 0.52	& 0.53	& 0.85	& 0.81	& 0.08	& 0.88	& 0.77	& 0.78	& 0.93	& 0.9	& 0.06 \\ \hline
Accuracy         	& 0.54	& 0.5	& 0.5	& 0.91	& 0.85	& 0.04	& 0.88	& 0.73	& 0.75	& 0.95	& 0.92	& 0.04 \\ \hline
Maintainability  	& 0.84	& 0.64	& 0.67	& 0.91	& 0.86	& 0.03	& 0.98	& 0.91	& 0.92	& 0.97	& 0.95	& 0.02 \\ \hline
Trust            	& 0.48	& 0.49	& 0.48	& 0.92	& 0.87	& 0.02	& 0.61	& 0.6	& 0.6	& 0.96	& 0.92	& 0.03 \\ \hline
Accessibility    	& 0.54	& 0.54	& 0.54	& 0.93	& 0.89	& 0.009	& 0.98	& 0.98	& 0.98	& 0.99	& 0.96	& 0.004 \\ \hline
Other            	& 0.48	& 0.49	& 0.48	& 0.92	& 0.87	& 0.02	& 0.98	& 0.75	& 0.78	& 0.98	& 0.93	& 0.01 \\ \hline

\textbf{Average} & 0.63 & 0.54 & 0.55 & 0.87 & 0.82 & 0.07 & \textbf{0.87} & \textbf{0.78} & \textbf{0.79} & \textbf{0.93} & \textbf{0.89} & \textbf{0.06}  \\ \hline

\end{tabular*}
\end{table*}

\subsection{Classification results}
The classification results, in terms of the different performance measures, are shown in Table~\ref{Tab:results}. Our results show that VSM can be heavily disadvantaged by the vocabulary mismatch problem of privacy policies' text. In general, $BR_{VSM}$ performed poorly, with an average $F_2$ = 0.55. In comparison, the word embeddings based approach GloVe did better in classifying all types of NFRs ($F_2 = 0.79$, $HL = 0.06$). This can be explained based on the fact that VSM vectorizes policy statements based on the TF.IDF scores of their individual words. In comparison, word embeddings capture the semantic meaning of words during vectorization. For example, words such as \textit{``improve''} and \textit{``enhance''}, which commonly appear in privacy policies, are not related according to VSM. However, GloVe assigns a similarity score of 86\% to these two words.   

We also observe that the GloVe based approach achieved its best performance when classifying accessibility, legal, and maintainability statements. One potential reason is that, in our dataset of policy statements, the overlap between these categories is low. For example, none of the accessibility statements overlap with other labels. Legal and maintainability statements were also mapped into distinct categories, with overlap rates of 16\% and 0.091\%.

\subsection{Annotated Policy Presentation}
In the fourth and final step of our approach, we annotate the privacy policies in our dataset based on the classification of their data collection claims. Annotations are commonly used in academia and practice to convey additional information about artifacts~\cite{Glover07}. The value of annotation for learning has long been established in the literature~\cite{Marshall97}. According to O'Donnell,~\cite{Donnell04} \textit{``Annotating a text can be a powerful strategy to comprehend difficult material and encourage active reading''.} Nowadays, annotations are used to enhance users' comprehension of online text~\cite{Rau04,Glover07}.  

\begin{figure}\centering
\includegraphics[trim=9.5cm 0.0cm 9.5cm 0.1cm, clip=true,scale=0.65]{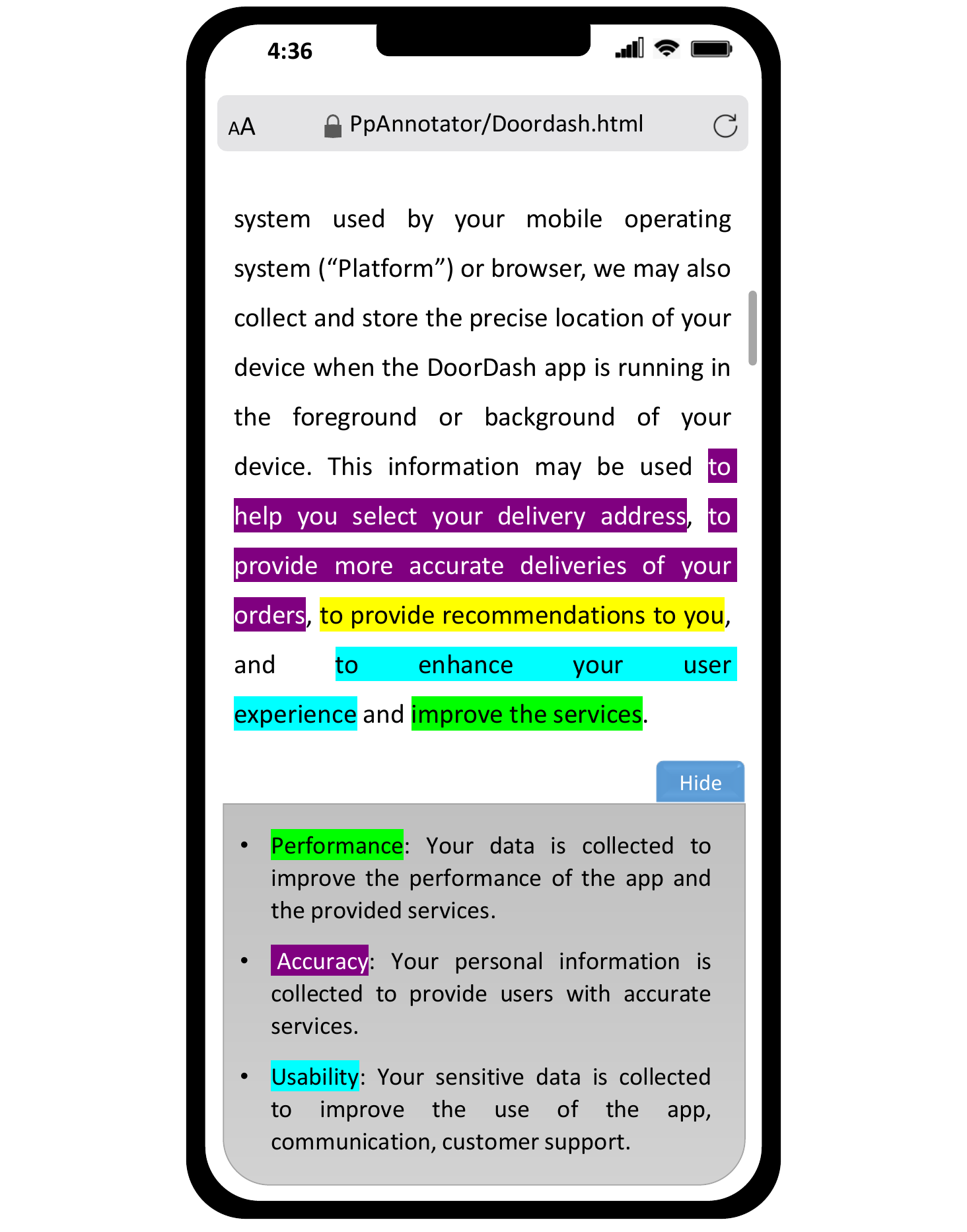}
\caption{A screen-shot of the annotated privacy policy of DoorDash.}
\label{fig:html}
\end{figure}

Annotations can be textual, such as comments or tags which add additional information to the document, or graphical, such as images, underlining, and coloring which help the reader identify important parts of the document with a minimum cognitive overhead. In our approach, we utilize hypertext annotations to generate more accessible DSE privacy policies. In particular, we created a JavaScript-enabled Web interface\footnote{\url{https://segroup-2318.github.io/PPannotator/DSE.html}} to display annotated policies to DSE users. A screenshot of our interface is shown in Fig.~\ref{fig:html}. The interface implements two types of annotations:

\begin{itemize}
\item Textual: For each classified data collection statement in each policy, we display a comment that explains what type of quality attribute, or NFR, the statement refers to.  

\item Graphical: We color-code the different classified statements for easier access. An optional key panel which elaborates more on the different types of annotations is also provided.
\end{itemize}      
In the next section, we evaluate the effectiveness of these two types of annotations on the comprehensibility of the privacy policies of DSE apps.

\section{Evaluation}
\label{sec:eval}
In the first phase of our analysis, we automatically annotated the data collection claims in the privacy policies of a large number of popular DSE apps. In this phase, we evaluate the value of such annotations for the end-users of DSE apps. In particular, to answer our second research question (\textit{RQ$_2$}), we conducted a human experiment with 18 active DSE users. Our objective is to measure the impact of annotating DSE privacy policies (by highlighting and color-coding their data collection practices) on their comprehensibility. 

\subsection{Study Participants}
We recruited 18 DSE users using convenience sampling to participate in our study. All subjects were college students. To participate in our experiment, we asked our subjects about the specific DSE platforms they have been actively using. Only students who have used DSE apps at least once during the 30 days prior to the experiment were allowed to participate. All of our subjects declared that they have used Lyft, Airbnb, and DoorDash at least once before. Therefore, these apps were chosen to be our experimental subject apps. All subjects declared that they have never read the privacy policies of any of the DSE apps they have used before. This confirmation was necessary to ensure that our subjects did not have any prior knowledge that might influence their answers during the experiment. In total, our subject sample consisted of seven (N = 7) female and 11 (N = 11) male participants with an average age of 20 years old. Each subject was compensated with \$24 for participating in the study.  

\subsection{Procedure}
Assessing the comprehensibility of text artifacts is not a straightforward process~\cite{Wilson16Crwod,Wilson18}. In general, a measure which assesses the level of engagement with the text artifact can provide a reasonable indication of how comprehensible the artifact is. In the literature, the comprehensibility of software privacy policies is commonly assessed through a mixture of quantitative and qualitative measures, which can range from questionnaires to more complex metrics of comprehensibility~\cite{Miller12,Redmiles20}.

In our experiment, subjects were divided into three groups of six subjects each (N = 6). Each group was then randomly assigned to one of our three subject apps. We asked the group assigned to each app to navigate to the privacy policy of their app using a link we provided. For each app, we prepared two privacy policies: the original policy of the app and an annotated copy. Half (N=3) of the group of subjects assigned to each app were then asked to navigate to the original policy (control group) while the other half were asked to navigate to the annotated policy. 

We asked our subjects to use their smartphones to navigate to the policies as all of them (N=18) declared that they only used DSE apps through their phones. Our annotation engine's interface was developed using JavaScript and HTML. We did our best to maintain the exact same format (font type, policy structure, and content) of the original policies. For the annotated policies, we added the color and text annotations. We further added an option to show and hide the help panel at the bottom (Fig.~\ref{fig:html}). Each subject then was asked to answer two open-ended questions about their assigned policy. These questions, shown in Table~\ref{Tab:questionnaire}, were formulated to be short and directly tackle data collection claims in each policy. In addition, all subjects had to answer one close-ended question: \textit{How would you rate the readability of this policy?} The subjects had to pick one of 4 answers: 1- Not readable, 2- Somewhat readable, 3- Readable, and 4- Very readable.  

It is important to point out that these questions were not intended to measure the quality of our subject policies, rather the level of engagement of our subjects with them. Furthermore, our subjects were not aware of the purpose of the experiment; they were not told that we were studying the impact of annotations on their interactions with the policies. No time limit was enforced to finish the experiment, but we kept track of time for each individual subject. In what follows, we discuss our results in greater detail. 

\begin{table*}
\footnotesize 
\centering
\renewcommand{\arraystretch}{1.2}
\caption{Our privacy policy comprehensibility questions for DoorDash, Airbnb, and Lyft.}
\label{Tab:questionnaire}
\smallskip 
\begin{tabular*}{38em}{l | l }
\Xhline{1.5\arrayrulewidth}
\textbf{App} & \textbf{Questions}  \\ 
 \Xhline{1.5\arrayrulewidth}

\multirow{2}{*}{DoorDash}
& Why does DoorDash permanently store your location information? \\ 
%\cline{2-2}
& Why does DoorDash collect your site usage information?  \\	 
\Xhline{1.5\arrayrulewidth}

\multirow{2}{*}{Airbnb}
& Why does Airbnb allow hosts to share guest information with third parties?  \\ 

& Why would Airbnb disclose your information to court?   \\	 
\Xhline{1.5\arrayrulewidth}

\multirow{2}{*}{Lyft}
& Why would Lyft share your pictures with other riders?   \\ 

& What types of information does Lyft ask for to ensure your safety?  \\	 

\Xhline{1.5\arrayrulewidth}
\end{tabular*}
\end{table*}

%do the time chart 
%do the readability chart 
%Confirm consistency with text 
%Prrof read and add citations where you think necessary

\subsection{Results}
We start our results analysis by grading the open-ended questions in our questionnaire. We used a three-point system for grading our subjects answers (0 = wrong, 1 =  partially correct, and 2 = correct). An answer is considered correct if it explicitly identifies the justification behind the data collection claim, a partially correct answer only includes an implicit statement related to the justification behind the claim, and a wrong answer does not provide either. Table~\ref{Tab:grading} shows our grading scheme in action\footnote{Graded responses are available in our complementary data}. Two judges independently graded our subject's answers according to a predefined rubric and the grades were then discussed to resolve any conflicts. Grades are shown in Table~\ref{tab:grades}. 

\begin{table*}
\footnotesize 
\centering
\renewcommand{\arraystretch}{1.2}
\caption{Grading three answers for the question: \textit{Why does DoorDash permanently store your location information?}}
\label{Tab:grading}
\smallskip 
\begin{tabular*}{48em}{p{3cm} | c | p{8cm} }
\Xhline{1.5\arrayrulewidth}
\textbf{Answer} & \textbf{Grade} & \textbf{Grade justification (rubric)}  \\ 
 \Xhline{1.5\arrayrulewidth}
To deliver my food &	1	& Partially correct answer which implicitly indicates that location data can be used for delivery \\
\hline
Tracking me	& 0	& Incorrect answer according to the policy \\
\hline
Enhance the accuracy of their GPS and recommend restaurants & 	2	& Correct answer as the policy indicates that DoorDash stores location information to \textit{“to provide more accurate deliveries of your orders”} and \textit{“to provide recommendations to you”}\\

\Xhline{1.5\arrayrulewidth}
\end{tabular*}
\end{table*}

The grades of the open-ended questions show that subjects who were exposed to the annotated policies scored on average higher on all questions (\textbf{Q$_1$} = 1.67/2 and \textbf{Q$_2$} = 1.44/2). In other words, they were more likely to either explicitly or implacability identify the rationale behind data collection claims in their policies. Subjects in the control group provided generic answers that were mostly incorrect (\textbf{Q$_1$} = 0.33/2 and \textbf{Q$_2$} = 0.44/2). Consider, for example, the first question about Airbnb's privacy policy: \textit{``Why does the app allow hosts to share guest information with third parties?''} Airbnb policy states that:

\begin{tcolorbox}[colback=gray!5,colframe=gray!40!black,]
Hosts may use third-party services to help manage or deliver their services, such as cleaning services or lock providers. Hosts may use features on the Airbnb Platform to share information about the Guest (like check-in and check-out dates, Guest name, Guest phone number) with such third-party service providers.
\end{tcolorbox}

Subjects in the control group only provided generic statements that we suspect were triggered by the phrase \textit{"share information with third parties"}. This was reflected in their answers: \textit{``To run ads that match my preferences''},  \textit{``To sell me ads''}, and \textit{``For money''}. In contrast, subjects who were exposed to the annotated Airbnb's policy were able to provide more accurate answers: \textit{``Improve app usability and user experience''}, \textit{``Provide cleaning services''}, \textit{``Improve my overall experience such as cleaning.''} In other words, subjects with annotated policies were able to gauge the rationale behind hosts sharing their guests' information with third parties rather than suggesting a generic reason.  

\begin{figure}\centering
\includegraphics[trim=0.5cm 4cm 1.5cm 1.5cm, clip=true,scale=0.4]{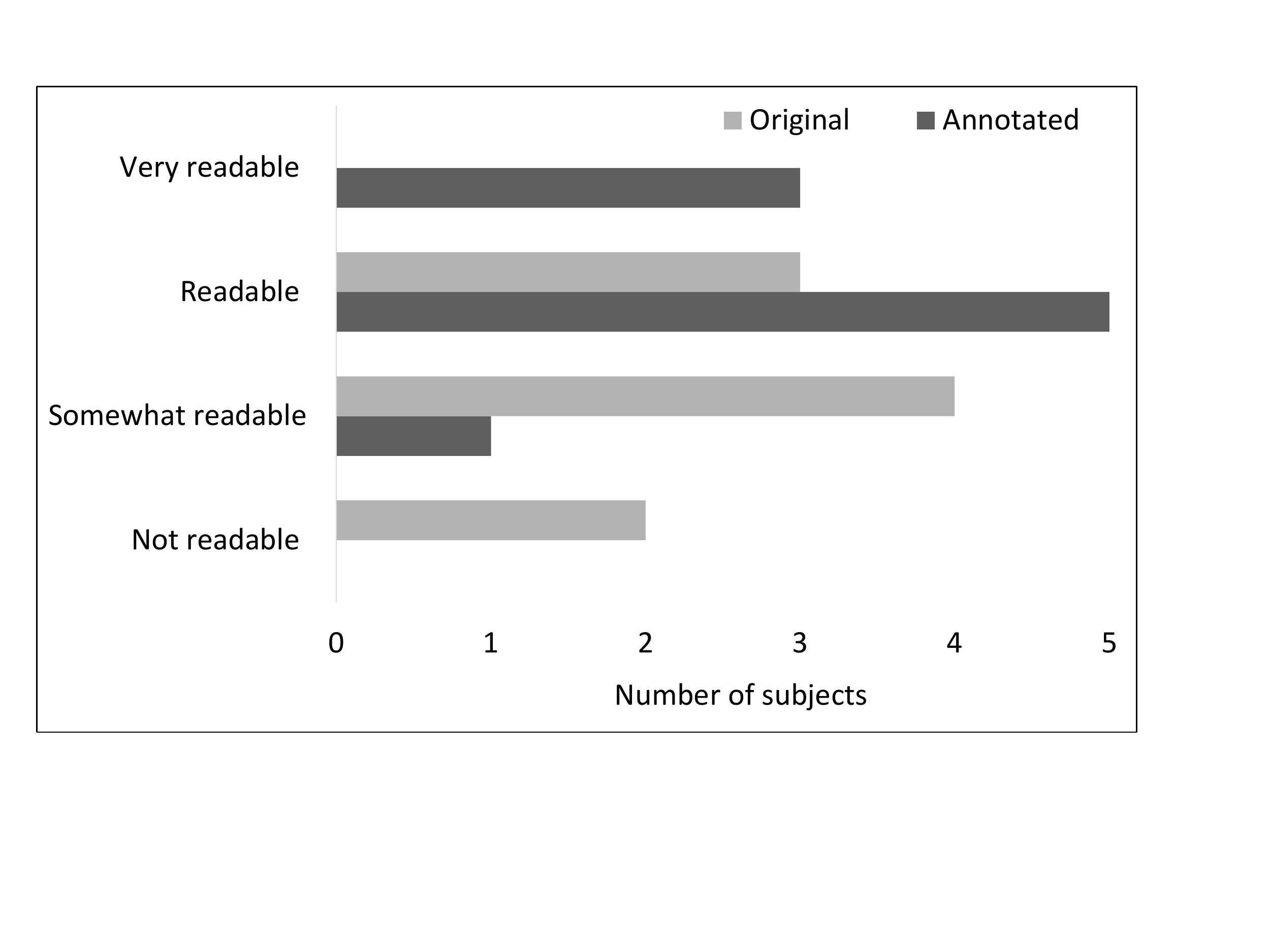}
\caption{Readability of our annotated and original privacy policies according to our study participants.}
\label{fig:readability}
\end{figure}

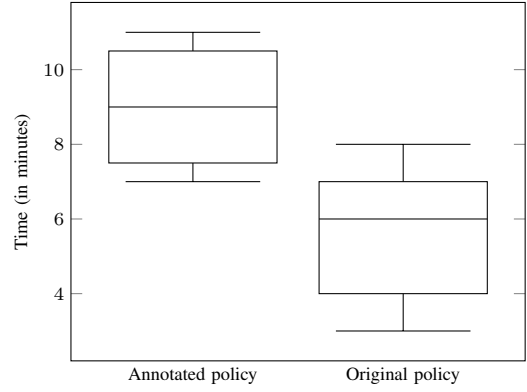
\begin{figure}[t!]
\centering
\scriptsize
\begin{tikzpicture}[scale=1]
\begin{axis}[
boxplot/draw direction=y,
ylabel={Time (in minutes)},
major x tick style=transparent,
xtick={1,2},
xticklabel style={text width=3cm,align=center},
xticklabels={Annotated policy, Original policy},
width=3 in,
height=2.5 in,
ylabel near ticks,
]
\addplot[boxplot prepared={
lower whisker=7, lower quartile=7.5,
median=9, upper quartile=10.5,
upper whisker=11},
draw=black
] coordinates{};

\addplot+[boxplot prepared={
lower whisker=3, lower quartile=4,
median=6, upper quartile=7,
upper whisker=8},
draw=black
] coordinates{};
\end{axis}
\end{tikzpicture}
\caption{The time our subjects spent on studying the privacy policy of their subject app and answering the questionnaire.}
\label{fig:time}
\end{figure}

%locate answers faster scrolling back and forth.
With regard to the third close-ended question, our results, shown in Fig.~\ref{fig:readability}, suggest that subjects who were exposed to the annotated policies constantly rated these policies to be more readable (average = 3.2) than subjects in the control group (average = 2.1). In related literature, the perceived readability of privacy policies has been found to lead users to actually read them~\cite{Ermakova16}. This effect was actually reflected in the time our subjects spent answering the questions. Specifically, our results (shown in Fig.~\ref{fig:time}) show that subjects who had access to the annotated policies took on average 10.4 minutes to go through our questions, while subjects who only had access to the original policies took on average 5.5 minutes to hand back their answer sheets. This effect was also reflected in the average length of answers provided by our subjects. Subjects who were exposed to our annotated policies provided longer answers (8 words on average) while subjects who were provided with the original policies provided brief short answers (4 words on average). For example, in answering the first question about DoorDash's privacy policy, answers of subjects who were exposed to the annotated policy ranged from 9 to 15 words: \textit{``The app keeps track of my location information to help their drivers find me faster''}, \textit{``So that they can find my address more accurately and make recommendations''},  and \textit{``Enhance the accuracy of their GPS and recommend restaurants.''}  While answers of subjects in the control group ranged from one to five words only: \textit{``To deliver my food''}, \textit{``Ads''}, \textit{``To know where I live.''}. 

In general, our results provide evidence that annotated policies encouraged longer engagement with the policy text. Our subjects were spending more time looking into the annotations key to provide their answers, while subjects in the control group seemed to be overwhelmed by the length of the policies and the lack of guidance on where to find answers. To our surprise, none of the subjects in either group actually used the search functionality available on their mobile phone's web browsers to locate answers, despite the fact that they were not told to do so. These results provide more evidence in favor of previous findings which raised major concerns about the accessibility and readability of privacy policies when displayed on smartphone displays. 

%Based on these results, we can conclude that using annotations based on NFRs helped to better communicate the rationale behind data collection claims in the privacy policies better with our subjects, furthermore, helped DSE app users to be more engaged with the policies and even think through suggestions that may improve the policy even more.   

\begin{table}
\footnotesize
\centering
\renewcommand{\arraystretch}{1.2}
\caption{Grades assigned to our subject's answers (0 = Incorrect, 1 =  partially correct, 2 = correct). S{1-3} used the annotated policy of DoorDash and S{4-6} used the original policy and so on. }
\label{tab:grades}
\smallskip
\begin{tabular}{l | c | c | c | c | c | c }
\Xhline{3\arrayrulewidth}

 \multirow{2}{*}{\textbf{Domain}} & \multicolumn{3}{c|}{\textbf{Annotated}} & \multicolumn{3}{c}{\textbf{Original}}    \\ \cline{2-7}
 
& \textbf{Subject} & \textbf{Q1} & \textbf{Q2} & \textbf{Subject} & \textbf{Q1} & \textbf{Q2} \\ \Xhline{3\arrayrulewidth}

\multirow{3}{*}{DoorDash}
& S1 & 1 & 1 & S4 & 1 & 0\\  \cline{2-7}
& S2 & 2 & 0 & S5 & 0 & 0\\  \cline{2-7}
& S3 & 2 & 2 & S6 & 1 & 1\\ \Xhline{3\arrayrulewidth}
 
%\multirow{3}{*}{Airbnb}
% & \cellcolor[gray]{0.8} S7 & \cellcolor[gray]{0.8} 2 & \cellcolor[gray]{0.8} 2 & \cellcolor[gray]{0.8} S10 & \cellcolor[gray]{0.8} 0 & \cellcolor[gray]{0.8} 2\\  \cline{2-7}
%& \cellcolor[gray]{0.8} S8 & \cellcolor[gray]{0.8} 1 & \cellcolor[gray]{0.8} 2 & \cellcolor[gray]{0.8} S11 & \cellcolor[gray]{0.8} 0 & \cellcolor[gray]{0.8} 0\\  \cline{2-7}
%& \cellcolor[gray]{0.8} S9 & \cellcolor[gray]{0.8} 2 & \cellcolor[gray]{0.8} 1 & \cellcolor[gray]{0.8} S12 & \cellcolor[gray]{0.8} 0 & \cellcolor[gray]{0.8}0\\ \Xhline{3\arrayrulewidth}

\multirow{3}{*}{Airbnb}
 &  S7 & 2 &  2 &  S10 & 0 &  2\\  \cline{2-7}
&  S8 &  1 &  2 &  S11 & 0 & 0\\  \cline{2-7}
& S9 & 2 &  1 &  S12 & 0 & 0\\ \Xhline{3\arrayrulewidth}

\multirow{3}{*}{Lyft}
& S13 & 2 & 2 & S16 & 0 & 0\\  \cline{2-7}
& S14 & 1 & 2 & S17 & 1 & 1\\  \cline{2-7}
& S15 & 2 & 1 & S18 & 0 & 0\\ \cline{2-7}

& \cellcolor[gray]{0.6} Average & \cellcolor[gray]{0.6} 1.67 & \cellcolor[gray]{0.6} 1.44 &  \cellcolor[gray]{0.6} Average & \cellcolor[gray]{0.6} 0.33 & \cellcolor[gray]{0.6} 0.44 \\
\Xhline{3\arrayrulewidth}
\end{tabular}
\end{table}

\section{Discussion and Impact}
\label{sec:discuss}
As of 2022, privacy remains one of the most pressing issues in software development~\cite{Hadar18}. Consumer data is being constantly collected, aggregated, and exploited for running targeted ads or even sold to third-party agencies~\cite{Benbenisty21}. Existing research on privacy policy is concentrated on detecting policy violations through static code analysis as well as investigating the structure and completeness of such policies~\cite{Ebrahimi20,Bhatia18,Wang18,Bhatia19}. In this paper, we argue that privacy is better understood at a domain level. In other words, each application domain (e.g., health~\cite{Haggag21,Peyton07}, social media~\cite{Kaplan21}, online dating~\cite{Cobb17}, etc.) has its own privacy paradox. Generic solutions, while might help understand the grand challenges of privacy, they often miss the contextual characteristics of specific operational environments. The problem becomes even more challenging in a domain such as the Sharing Economy, where privacy concerns can extend to the physical privacy of users.

Our work in this paper is a step toward helping users of the Sharing Economy apps to better understand the privacy practices of their apps, thus, making more rational socio-economic decisions. In particular, in the Sharing Economy market, several apps are available for each type of service~\cite{Williams20,Tushev21}. Users often make their DSE app selection decisions based on factors such as the service fee or the quality of provided service while undermining issues such as privacy. Providing more comprehensible policies can help users to be more engaged with their app policies and better understand the privacy risks associated with using each app, and hopefully, use that information as another deciding factor when selecting which app to use or work for.

Our findings in this paper can also help DSE app developers to draft more comprehensible privacy policies and to better justify their data handling practices. A more comprehensible privacy policy not only encourages users to read the policy, but can also enhance their trust in the platform, thus their willingness to disclose more information~\cite{Bansal08}. Trust has been found to play a major role in DSE app survival. In particular, users need to have a certain level of trust in the platform before a transaction can take place~\cite{Ma17,Ert16}. Drafting more readable policies can directly and positively enhance this trust~\cite{Ermakova16}. Existing policy evaluation protocols can also start considering annotations as a measure of policy quality~\cite{Miller12,Ryker02}. In particular, in addition to standard readability measures~\cite{Fabian17,Miller12,Flesch49}, the presence of informative annotations can be considered an indication of a high-quality policy. 

Existing evidence also suggests that privacy policies become even more inaccessible to the average Internet user when displayed on mobile phones~\cite{Singh11}. In general, mobile displays are not convenient for reading and comprehending unfamiliar technical and legal materials as privacy policies. In the Sharing Economy, most apps redirect their users to a web-based policy rather than crafting a different format policy for the mobile version. Our work in this paper has shown that such problems can be mitigated by using mobile-enabled annotations.    

In terms of limitations, concerns can be raised about the potential information overload that might stem from using annotations. In particular, privacy policies are already long and hard to read~\cite{Jensen05,Fabian17,Kaplan21}, adding more information might make them even less appealing for some users to engage with. In our approach, we restricted our annotation to simple text and colors, thus we expect such effect to be minimal. However, we acknowledge the fact that certain forms of textual or visual annotations might be more informative or less cognitive heavy than others. Further experimentation would be necessary to uncover such forms. Other concerns might be raised about altering the legally-binding text of privacy policies with unvetted text. This issue can be resolved with the help of legal experts who can craft the appropriate wording for different annotation statements.  

\section{Threats to Validity}
\label{sec:threats}
Our study has several limitations that could potentially limit the validity of the results. In terms of internal validity, threats could stem from the fact that human judgment was used to annotate the privacy policies in our dataset. Different judges might annotate data collection claims differently. Despite these concerns, it is common in text classification tasks to use humans' judgment to prepare the ground-truth. Therefore, these threats are inevitable; however, they can be partially mitigated by using multiple judges and a systematic coding and conflict resolution process. Another threat might stem from the specific classification configurations used in our analysis. For instance, we only experimented with SVM, using VSM and GloVe to generate our feature vectors. Other classification configurations might arrive at different results.  

Internal validity threats might stem from the fact that only college students were used as subjects in our human evaluation study. The decision to use college students was enforced to make sure that our subjects were all at the same educational level, thus eliminate any confounding factors related to our subjects' literacy level. However, we acknowledge the fact that further experimentation with different groups of users (age, gender, educational level, employment status, etc.) is necessary to confirm our findings. Another threat might stem from the fact that subjects' responses were graded by only two experts. We tried to minimize this threat by using a simple rubric and a three-point scale for grading. We also make our data publicly available to enable independent replications of our study.   

A threat to external validity stems from the fact that we only experimented with 30 DSE apps for classification. Other apps might have different format policies. However, as mentioned earlier, the apps we included in our analysis were selected as they were the most popular in their specific categories. Privacy concerns are more likely to manifest over these popular apps than smaller, less well-known apps. Other threats can originate from our human experiment. Only 18 subjects participated in the study. Therefore, the results might not necessarily be generalizable beyond this group. Nonetheless, we believe that our experiment was sufficient to provide preliminary evidence on the value of annotations. 

\section{Conclusion}
\label{sec:conclusion}
In this paper, we proposed an automated approach for annotating the privacy policies of Sharing Economy apps. Our approach uses text classification to map individual data collection claims in privacy policies to different categories of non-functional requirements. Our objective was to make these policies more comprehensible to the average DSE user. Our results showed that SVM trained over word embeddings of privacy policies text can produce accurate classification results. The comprehensibility of the annotated policies was then evaluated through a user study with 18 DSE users. The results showed that textual and visual annotations helped our study participants to better understand the rationale behind the data collection practices of their DSE apps. 

In terms of future work, the study in this paper will be extended with more experimentation. In particular, we will continue to validate our annotated policies with more subjects. Our goal is to determine the most effective types of textual and visual annotations for each specific DSE domain. As mentioned earlier, privacy requirements and goals vary greatly among different domains of DSE (eg., lodging, ridesharing, freelancing, etc.). Therefore, it is important to conduct studies at these different domains to better understand what works (or does not work) for different domains.

%\section*{Acknowledgment}
%A placeholder for acknowledging the funding agency. A placeholder for acknowledging the funding agency.   

\bibliographystyle{IEEEtran}
\bibliography{RE}

\end{document}